\RequirePackage[2020-02-02]{latexrelease}
\documentclass[twocolumn,english,aps,prl,10pt,superscriptaddress]{revtex4}
\usepackage[T1]{fontenc}
\usepackage{amsmath,graphicx,amssymb,epsfig,babel,dsfont}
\usepackage{mathrsfs}
\usepackage{color}

\def\bbm[#1]{\mbox{\boldmath $#1$}}

\begin{document}

\title{Smart windows passively driven by greenhouse effect}

\author{G. Boudan}
\affiliation{Laboratoire Charles Fabry, UMR 8501, Institut d'Optique, CNRS, Universit\'{e} Paris-Saclay, 2 Avenue Augustin Fresnel, 91127 Palaiseau Cedex, France.}
\affiliation{Thales Research and Technology France, 1, Avenue Augustin Fresnel, F-91767 Palaiseau, Cedex France.}
\author{E. Eustache}
\affiliation{Thales Research and Technology France, 1, Avenue Augustin Fresnel, F-91767 Palaiseau, Cedex France.}
\author{P. Garabedian}
\affiliation{Thales Research and Technology France, 1, Avenue Augustin Fresnel, F-91767 Palaiseau, Cedex France.}
\author{R. Messina}
\affiliation{Laboratoire Charles Fabry, UMR 8501, Institut d'Optique, CNRS, Universit\'{e} Paris-Saclay, 2 Avenue Augustin Fresnel, 91127 Palaiseau Cedex, France.}
\author{P. Ben-Abdallah}
\email{pba@institutoptique.fr}
\affiliation{Laboratoire Charles Fabry, UMR 8501, Institut d'Optique, CNRS, Universit\'{e} Paris-Saclay, 2 Avenue Augustin Fresnel, 91127 Palaiseau Cedex, France.}

\date{\today}

\begin{abstract}
The rational thermal management of buildings is of major importance for the reduction of the overall primary energy consumption.  Smart windows are promising systems which could save a significant part of this energy. Here we introduce a double glazing system made with a thermochromic metal-insulator transition material and a glass layer separated by an air gap which is able to switch from its insulating to its conducting phase thanks to the greenhouse effect occuring in the separation gap.  We also show that this passive system can reduce  the incoming heat flux by $30\%$ in comparison with a traditional double glazing while maintaining the transmittance around $0.35$ over $75\%$ of visible spectrum.
\end{abstract}
\maketitle

Nowadays buildings consume about $40\%$ of the primary energy used in the world~\cite{Ramesh}. Therefore their thermal management is a major concern in the context of current global challenge of energy efficiency. Smart windows could play a major role to respond to this challenge. These windows can be used to control heat flux exchanged between the indoor and external environnment. 
One way to achieve this control is by tuning the radiative properties of these systems.  Hence, by controlling  the overall emissivity/transmittivity of a window surface it is possible to facilitate or reduce heat exchange with the environment and therefore to act directly on the indoor temperature. Several strategies have been proposed so far to perform such a control. For example, single glazings made of thermochromic materials have been suggested~\cite{Kim,Zheng,Chang} to tailor the optical properties of windows with respect to their temperature. Among these materials vanadium dioxide (VO$_2$) a metal-insulator transition (MIT) material has been largely investigated. Due to their pronounced change of optical properties in the infrared these phase-change materials have found numerous applications in the fields of thermal management~\cite{pba2013,pba2014,pba2014_bis,Reddy2018,Kats}, energy storage~\cite{Kubytskyi} and information treatment~\cite{pba2016}. However this material  undergoes a reversible structural phase transition at relatively high temperature ($T_C \sim 68^{\circ}$C) making its employment challenging in smart windows. Moreover, to date, there is no real alternative candidate to this material presenting a pronounced MIT close to the ambient temperature. 
Electrochromic materials have also been proposed~\cite{Cannavale,Chen} to modify the optical properties of smart windows using an external bias voltage. Hence, by using VO$_2$ solid-electrolyte films, the light transmittance of such gate-controlled smart windows can be dramatically modified by tuning the hydrogen-ion doping through gating voltage. However, although this optical control can operate at ambient temperature it requires an external energy supply. 
Moreover, electrochromic windows are mainly effective in blocking visible light and not  infrared radiation so that their thermal performances remain today very weak.

In the present letter we introduce a passive smart window based on a double glazing system (Fig.~1) realized with a VO$_2$ film deposited on the internal side of the glazing. The two panes of this glazing are separated by a few-milimeter thick gap filled with noble gas.

\begin{figure}[bt]
		\includegraphics[angle=0,scale=0.34]{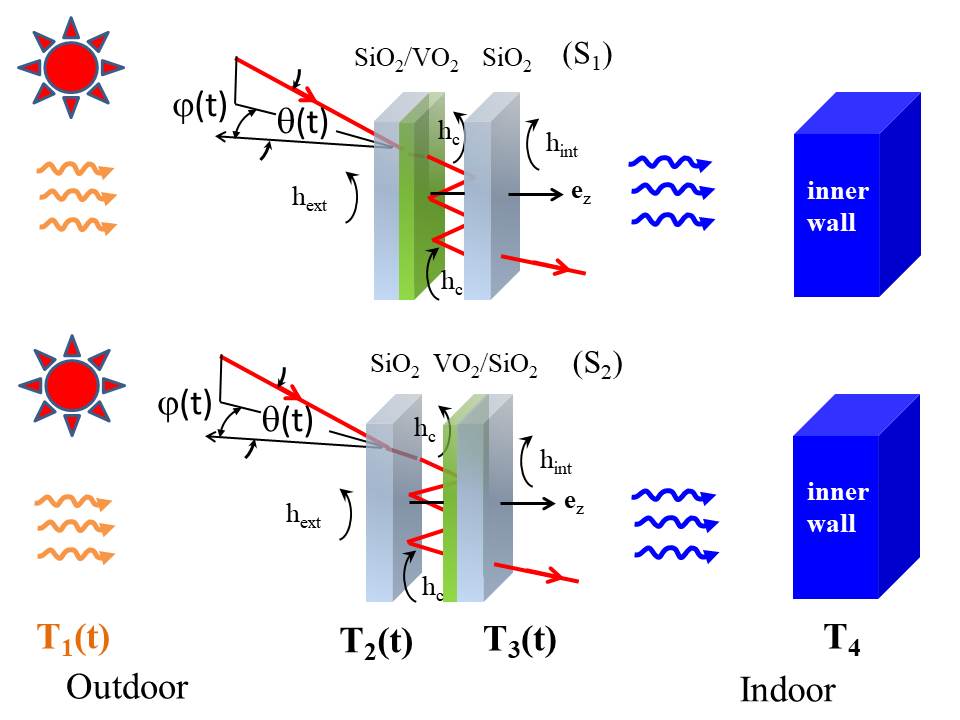}
		\caption{Sketch of the passive smart window made with a double glazing of two SiO$_2$ layers separated by a gap filled of noble gas and a VO$_2$ thin film deposited on the inner side of one of the two glass panes. The temperature of the VO$_2$ layer is driven by the greenhouse effect occuring between the two panes thanks to the external sunshine in the visible range, by the heat flux radiated in the infrared by the surrounding environnment and the inner wall and by the convective exchanges.}
		\label{Fig:Syst}
\end{figure}

 Following the evolution during the day of external conditions of sunshine evolve during the day the temperature of this MIT material changes thanks to the greenhouse effect occuring between the two panes. By optimizing the geometry of this system we show that we can at the same time control its transmittance in the visible range and the incoming heat flux. More specifically, by analyzing the temporal evolution of this thermo-optical system driven by the external sunshine and the heat flux radiated in the infrared by the external environment and the inner walls we show that our smart-window design results in a significant reduction of the heat flux reaching the inner side, compared to traditional windows, in conjunction with a low reduction of the transmittance in the visible range of the spectrum.

To start let us describe the physical characteristics of system. The active part is made of a vanadium dioxide (VO$_2$) film which 
undergoes a first-order transition (Mott transition~\cite{Mott}) from a high-temperature metallic phase to a low-temperature 
insulating phase~\cite{Barker,Barker2} at $T_{C}$. This film is deposited on a glass layer which is partially transparent in the infrared range~\cite{Palik}. As a result of
the phase transition both the emissivity and the reflectivity of this bilayer change radically  (Fig.~2) on its two sides both in the visible and in the infrared. The outside pane of the window is assumed to be in contact with a thermal bath having temperature $T_{1}(t)$ and is illuminated by the sun with a time-varying flux $\Phi_2(t)$ depending on the sun trajectory during the day. For the sake of simplicity, the field radiated by the bath can be assimilated to the field radiated by a blackbody at the same temperature. On its opposite side the window interacts with the inner walls of average emisssivity $\epsilon$ at temperature $T_{4}$. Due to the thermal inertia of walls,  $T_{4}$ is assumed here constant during a solar cycle.

\begin{figure}[bt]
		\includegraphics[angle=0,scale=0.38]{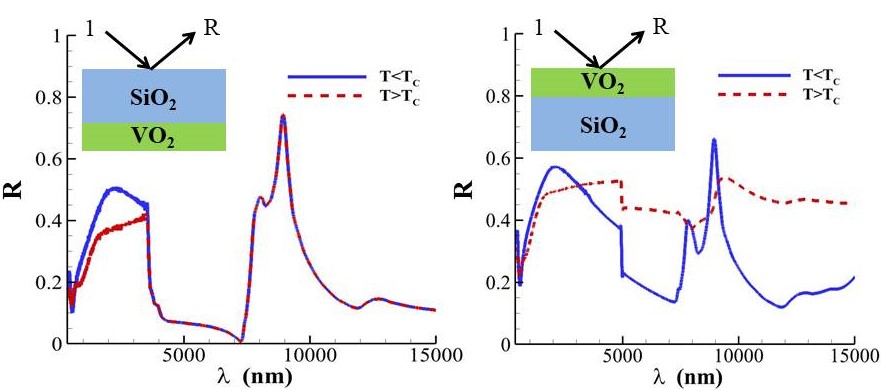}
		\caption{Reflectance spectra (averaged over the two polarization states) of SiO$_2$-VO$_2$ (left) and VO$_2$-SiO$_2$ (right) bilayer across the MIT process.  The glass layer is $5\,$mm thick while the  thickness of VO$_2$ film is $100\,$nm.}
		\label{reflexion}
\end{figure}
The optical properties (emissivity/transmittivity) of the window are driven by the temperature of each pane and therefore by the heat flux they exchange between them and with the outdoor and indoor environments. The time evolution of the thermal state $(T_2(t),T_3(t))$ of the window under the external driving is governed by the coupled energy-balance equations 
\begin{equation}
\begin{split}
	C_i\frac{d{T}_i}{dt}&=\mathscr{P}_ i(T_1,...,T_4;t)+\Phi_i(t)\\
&\,+\mathscr{P}_{\text{conv}\rightarrow i}(T_{i-1},T_i,T_{i+1};t) ,\quad i={2,3}.
	\label{Eq:dynamic}
\end{split}
\end{equation}
Here $\mathscr{P}_i$ denotes the total surfacic power received by the $i^{th}$ element from all other elements of the system, $\Phi_i$ the contribution of solar flux, $\mathscr{P}_{\text{conv}\rightarrow i}$ the losses/gains due to the convective exchanges and $C_i$  the heat capacity per unit surface for the pane $i$ of the window. The power exchanged per unit surface by each plate ($i=2,3$) with its surrounding by convection is calculated from the convective heat transfer coefficients $h_c$, $h_\text{ext}$ and $h_\text{int}$ (within the cavity formed by the two panes, between the external medium and the external pane and between the inner medium and the internal pane) using the linear Newton's law of cooling 
\begin{equation}
\mathscr{P}_{\text{conv}\rightarrow 2,3} =h_{c}(T_{3,2}-T_{2,3})+h_\text{ext,int}(T_{1,4}-T_{2,3}),
	\label{Natural convection}
\end{equation}
which takes into account the convective interactions on all sides of the layer. We consider here a double glazing with a cavity filled with a noble gas, such as Argon, and we assume each pane cooled or heated up by  natural convection on its opposite side.  With vertical windows and  temperature differences between the panel and its surrounding smaller than $\Delta T\sim50^\circ$C the corresponding heat transfer coefficients can be reasonably  set to $h_c=1\,$W$\cdot$m$^{-2}\cdot$K$^{-1}$ and  $h_\text{ext,int}=5\,$W$\cdot$m$^{-2}\cdot$K$^{-1}$ which correspond to the typical values for vertical panels in these conditions~\cite{Awbi,McAdams}. Hence, the power exchanged  by convection between each panel and its surrounding is about $\mathscr{P}_{\text{conv}\rightarrow 2,3}\sim(h_{c}+h_\text{ext})\Delta T$ that is a third of the solar irradiation. In this preliminary study, we do not include the heat exchanges due to forced convection or turbulent exchanges between each pane and its surrounding environment. These effects as well geometric effects such as the window inclination and sizing will be analyzed in future works. The heat flux across any plane parallel to the interacting surfaces is given by the normal component of Poynting vector
\begin{equation}
  \varphi(z)= \langle \mathbf{E}(z)\times\mathbf{H}(z)\rangle \cdot \boldsymbol{e}_z.
  \label{Eq:flux}
\end{equation}
Here $\boldsymbol{e}_z$ denotes the unit vector normal to the panes of windows, $\mathbf{E}$ and $\mathbf{H}$ are the 
local electric and magnetic fields which are generated by the randomly fluctuating source currents in each solid and by the fields radiated by media $1$ and $4$ while $\langle . \rangle$ represents the classical statistical averaging over all field realizations.
Using the Fourier decomposition of electric and magnetic fields $\varphi(z)$ can be recasted in terms 
of the field correlators $\mathfrak{C}_j^{\phi,\phi'}(\omega,\kappa)= \langle E_j^\phi(\omega,\kappa)E_j^{\phi'\dagger}(\omega,\kappa)\rangle$ of local field amplitudes in polarization $j$~\cite{RMP,Latella}
\begin{equation}
\begin{split}
     \varphi(z) = 2\epsilon_0c^2 \!\!\sum \limits_{\underset{\phi=\left\{ +,-\right\}}{j=s,p}}\int_{0}^{\infty}\!\!\frac{d\omega}{2\pi}\underset{\kappa<\omega/c}{\int}\!\!\! \kappa\frac{{\rm d} \kappa}{2 \pi}\frac{\phi k_z}{\omega} \mathfrak{C}_j^{\phi,\phi}(\omega,\kappa),
\label{Eq:base}
\end{split}
\end{equation}
where $ \kappa$ and $k_z=\sqrt{\frac{\omega^2}{c^2}-\kappa^2}$ denote the parallel and normal 
components of the wavector. The correlators $\mathfrak{C}_j^{++}$ and  $\mathfrak{C}_j^{--}$ in each region of the system can be written in term of correlators of fields radiated by each solid and by the external baths. We summarize below the calculation of one of such correlators (say $\mathfrak{C}_j^{++}$ in the windows cavity), the calculation of all others correlators being similar. 
The right and left propagating fields $E_c^{\pm}$ in this cavity are related to the fields in the region of baths  $E_{1,4}^{\pm}$ and to the fields $E_i^{\pm}$ ($i=2,3$) emitted by each pane by the following systems (by omitting to write the polarization for clarity reasons)
\begin{equation}
\begin{split}
E_c^{+}=E_2^{+}+\rho_2^{+}E_c^{-}+\tau_2^{+}E_1^{+},\\
E_c^{-}=E_3^{-}+\rho_3^{-}E_c^{+}+\tau_3^{-}E_4^{-},
\end{split}\label{Eq:cavity1}
\end{equation}
where $\rho_i^{\pm}$ and $\tau_i^{\pm}$ are the reflection and transmission operators of the layer $i$  toward the right ($+$) and the left ($-$), respectively. By solving this system with respect to the cavity field we get
\begin{equation}
\begin{split}
E_c^{+}&=U_{23}[E_2^{+}+\tau_2^{+}E_1^{+}+\rho_2^{+}(E_3^{-}+\tau_3^{-}E_4^{-})]\\
E_c^{-}&=U_{23}[\rho_3^{-}(E_2^{+}+\tau_2^{+}E_1^{+})+E_3^{-}+\tau_3^{-}E_4^{-}]
\end{split},\label{Eq:cavity3}
\end{equation}
where we have set $U_{23}=[1-\rho_2^{+}\rho_3^{-}]^{-1}$. It follows that the correlators ($++$)  into the cavity reads in term of correlators of fields radiated by each solid
\begin{equation}
\begin{split}
\mathfrak{C}_c^{+,+}&=\mid U_{23}\mid^2[\mathfrak{C}_2^{+,+}+\mid \rho_2^{+} \mid^2 \mathfrak{C}_3^{-,-}\\
&\,+ \mid \tau_2^{+} \mid^2 \mathfrak{C}_1^{+,+}+\mid \rho_2^{+} \mid^2 \mid \tau_3^{-} \mid^2 \mathfrak{C}_4^{-,-}].
\end{split},\label{Eq:cavity4}
\end{equation}
But, according to the fluctuation dissipation theorem~\cite{Rytov}, these correlators reads
\begin{equation}
\begin{split}
\mathfrak{C}_2^{+,+}&=\mathcal{S}(T_2)(1- \mid \rho_2^{+} \mid^2 -\mid \tau_2^{+} \mid^2 ),\\
\mathfrak{C}_3^{-,-}&=\mathcal{S}(T_3)(1- \mid \rho_3^{-} \mid^2 -\mid \tau_3^{-} \mid^2 ),\\
\mathfrak{C}_1^{+,+}&=\mathcal{S}(T_1),\\
\mathfrak{C}_4^{-,-}&=\mathcal{S}(T_4),\\
\end{split}\label{Eq:cavity5}
\end{equation}
with 
\begin{equation}
\mathcal{S}(T)=\pi\frac{\omega}{\epsilon_0 c^2}\frac{1}{k_z}N(\omega,T),
\end{equation}
and $N(\omega,T)=\frac{\hbar\omega}{2}\coth(\frac{\hbar\omega}{2k_B T})$. 
Finally, the radiative power received by the two panes from the rest of the system (except the sun) reads
\begin{equation}
\begin{split}
\mathscr{P}_2&=\varphi(0)-\varphi(-\delta_2),\\
\mathscr{P}_3&=\varphi(d+\delta_3)-\varphi(d),
\label{net power}
\end{split}
\end{equation}
$\delta_{2,3}$ being  the panes thickness.

\begin{figure}[bt]
		\includegraphics[angle=0,scale=0.39]{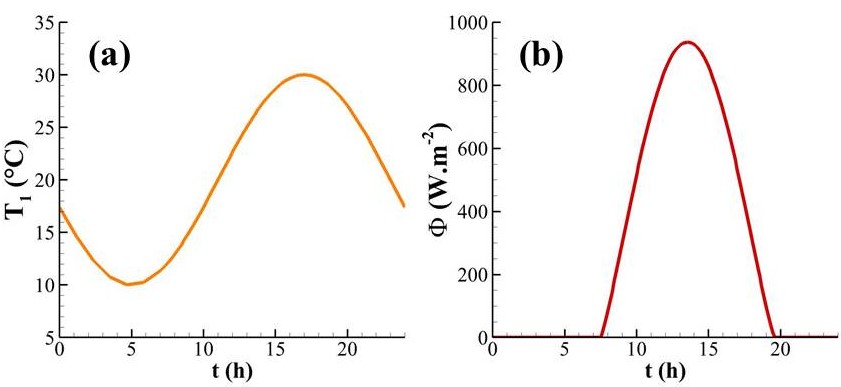}
		\caption{Time evolution during a typical summer day of the external temperature (a) and that of solar flux (b) at a lattitude of 48.2$^\circ$C.}
\label{sources}
\end{figure}
The solar flux plotted in Fig.~3(b) is calculated from the AM1.5 Global spectrum~\cite{NREL} over a daily cycle for a window facing south. We also assume a sinusoidal variation of the outdoor temperature [Fig.~3-(a)]  with a minimal value $T^\text{min}_1=10^{\circ}$C at $5\,$a.m. and a maximal value $T^\text{max}_1=30^{\circ}$C at at $4\,$p.m., corresponding to a typical variation of this temperature during summer. Finally, in order to demonstrate the operating mode of the window we assume the inner wall perfectly absorbing (i.e. $\epsilon=1$). The time evolution of these two primary sources being very slow in comparison with the thermalization time of glazing ($\sim$ few seconds to the minute), the energy balance Eq.~(\ref{Eq:dynamic}) can be solved in the adiabatic approximation.

The solutions $(T_2(t),T_3(t))$ of this balance equation corresponding to the external conditions (solar flux $\Phi(t)$, outdoor temperatures $T_1(t)$) given the indoor temperature $T_4=25^{\circ}$C are plotted in Fig.~4 for different thicknesses of VO$_2$ film and compared to the thermal state of a simple glass-glass double glazing. As shown in Figs.~4(a) and (b) the temperature of the active layer can reach, at the midday, levels significantly higher than the critical temperature $T_C$. The comparison of $T_1$ (resp.  $T_2$) with the temperatures of a double (glass-glass) glazing of same thickness demonstrates the presence of a strong greenhouse effect in the designed window due to the trapping of the infrared light in the separation gap between the two panes. The magnitude of this effect  is directly related to the thickness of VO$_2$ film. As shown in Figs.~4(a) and (b), the thicker is this film the greater is the blocked infrared radiation in the separation gap between the two panes of glazing. The heating induced by this greenhouse effect in the smart window surpasses the one traditionally observed in the traditional double-glazing window  by more than $100\%$ with VO$_2$ film $100\,$nm thick in $S_1$-type configuration and it is still $55\%$ higher with a film $10\,$nm in the same configuration. The physical consequences of this heating are shown in Figs.~4(c) and (d) demonstrating both the thermal and optical performances of this smart window. 
\begin{figure}[bt]
		\includegraphics[angle=0,scale=0.38]{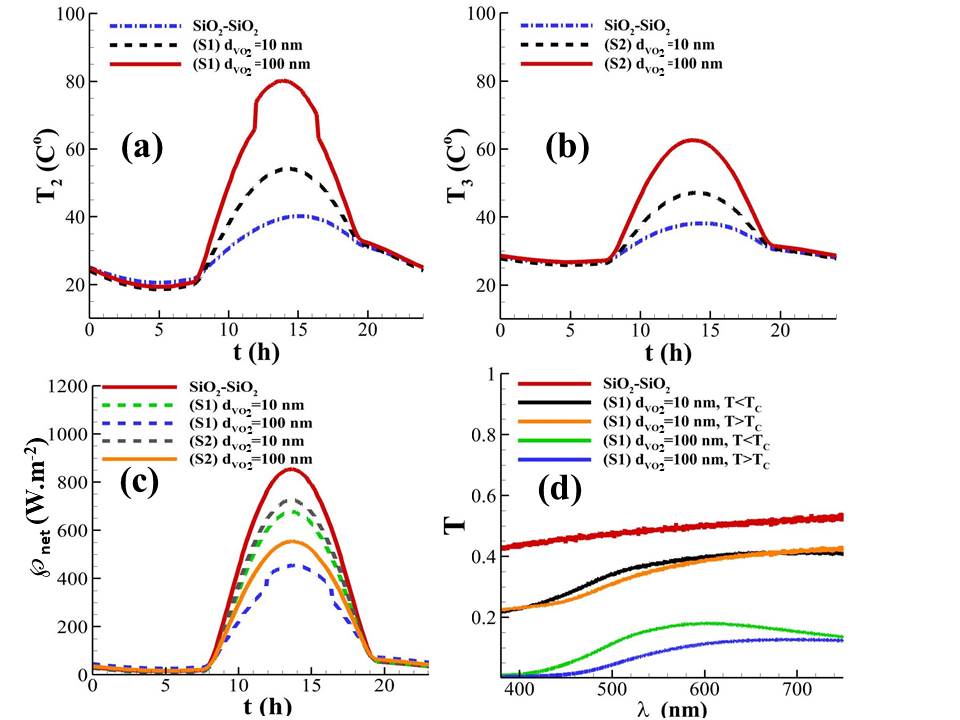}
		\caption{Thermal state and performances of the smart window. (a) Evolution of panes temperature in a $S_1$-type window. (b) Temperatures in a window of $S_2$-type. (c) Incoming net power $\mathscr{P}_{net}$ with respect to time for a $S_1$ and $S_2$ window and its comparison with a traditional double glazing. The glass and VO$_2$ films have thicknesses $d_{\text{SiO}_2}=5\,$mm and $d_{\text{VO}_2}=10\,$nm or $d_{\text{VO}_2}=100\,$nm, respectively. (d) Transmittance in the visible range of the smart window ($S_1$ configuration) below (resp. beyond) the critical temperature $T_C$ and its comparison with a double glazing SiO$_2$-SiO$_2$.}
\label{performances}
\end{figure}
The net incoming heat flux $\mathscr{P}_\text{net}$, with respect to time, plotted in Fig.~4(c) is drastically reduced in comparison with a traditional window (double glazing SiO$_2$-SiO$_2$). This reduction is approximately $400\,\text{W}\cdot\text{m}^{-2}$ at the midday when the VO$_2$ film $d_{\text{VO}_2}=100\,$nm is deposited on the inner side of the external pane ($S_1$ configuration in Fig.~1), a value representing $40\%$ of the solar irradiation at $48.2^{\circ}$ lattitude and about $55\%$ of the incoming flux through a traditional window. Such isolating behavior exceeds by far the performances of all previous smart windows~\cite{rewiew_Feng,rewiew_Cui}. On the other hand, when the MIT layer is on the opposite pane ($S_2$ configuration) the incoming net flux becomes, in comparison, more important, the temperature of this pane being increased by the greenhouse effect. For $10\,$nm thick VO$_2$ the thermal insulation is reduced by a third.

Beside the thermal characteristics of the window we show in Fig.~4(d) the transmittance 
\begin{equation}
T(\omega)=\int_{0}^{\pi/2}[t_s(\omega,\theta)+t_p(\omega,\theta)]\cos \theta \sin \theta d\theta
\label{transmittance}
\end{equation}
of the smart window in the visible range for both phases of thermochromic material. Here $t_s$ (resp. $t_p$) denotes the directionnal spectral transmission of the window under an incident angle $\theta$ (with respect to the normal, see Fig.~1)  in $s$ (resp. $p$) polarization. Notice that this transmittance depends weakly, in the visible, on the crystallographic state of VO$_2$ film, the main changes taking place mainly in the infrared. Although relatively small, this transmittance ($\sim 35\%$ for $10\,$nm thick VO$_2$) over $75\%$ of spectrum is comparable with that of usual double glazing SiO$_2$-SiO$_2$. This transmittance falls down to $20\%$ with $100\,$nm thick VO$_2$, making the window partially opaque. Nevertheless, this value could be significantly improved using a nanostructured MIT layer without significantly affecting its properties in the infrared, as discussed in Ref.~\cite{rewiew_Feng,rewiew_Cui,Taylor,Liu}. However, this could also significantly increase the overall fabrication cost. This problem as well as the optimization of the window will be addressed in a specific study.

In conclusion we have introduced a smart window based on a double glazing made with a SiO$_2$-VO$_2$  bilayer which is autonomously thermally and optically regulated by greenhouse effect. This system overcomes the primary weaknesses of thermochromic and electrochromic smart windows.
Beyond its application in the developpemnt of smart windows for more sustainable  buidlings, the passive insulation mechanism  introduced in the present work could find broader applications in the development of self-regulated insulating materials for complex systems. 

\section*{Acknowledgments}

This research was supported by the French Agence Nationale de la Recherche (ANR), under grant ANR-19-CE08-0034 (PassiveHEAT). P. B.-A. acknowledges discussions with E. Blandre and A. Losquin.

\section*{Data availability}

The data that support the findings of this study are available from the corresponding authors upon reasonable request.

\end{document}